\documentclass[letterpaper]{article} 
\usepackage{aaai24}  
\usepackage{times}  
\usepackage{helvet}  
\usepackage{courier}  
\usepackage[hyphens]{url}  
\usepackage{graphicx} 
\usepackage{natbib}  
\usepackage{caption} 
\usepackage{algorithm}
\usepackage{algorithmic}
\usepackage{longtable}
\usepackage{array}
\usepackage{newfloat}
\usepackage{listings}
\DeclareCaptionStyle{ruled}{labelfont=normalfont,labelsep=colon,strut=off} 
\floatstyle{ruled}
\newfloat{listing}{tb}{lst}{}
\floatname{listing}{Listing}
\title{The Problems with Proxies: Making Data Work Visible through Requester Practices}
\author{
    Annabel Rothschild\textsuperscript{\rm 1}, Ding Wang\textsuperscript{\rm 2}, Niveditha Jayakumar Vilvanathan\textsuperscript{\rm 1}, Lauren Wilcox\textsuperscript{\rm 1}, Carl DiSalvo\textsuperscript{\rm 1}, Betsy DiSalvo\textsuperscript{\rm 1}}
\affiliations{
    \textsuperscript{\rm 1}Georgia Institute of Technology\\
    85 5th St NW, Atlanta, Georgia 30332 USA\\
    \textsuperscript{\rm 2}Google Research\\
     1105 W Peachtree St NW, Atlanta, GA 30309\\
     
    \{arothschild, nvilvanathan3, cdisalvo\}@gatech.edu, drdw@google.com, lgw231@acm.org, bdisalvo@cc.gatech.edu\\

}

\usepackage{bibentry}

\begin{document}

\maketitle

\begin{abstract}
Fairness in AI and ML systems is increasingly linked to the proper treatment and recognition of data workers involved in training dataset development. Yet, those who collect and annotate the data, and thus have the most intimate knowledge of its development, are often excluded from critical discussions. This exclusion prevents data annotators, who are domain experts, from contributing effectively to dataset contextualization. Our investigation into the hiring and engagement practices of 52 data work requesters on platforms like Amazon Mechanical Turk reveals a gap: requesters frequently hold naive or unchallenged notions of worker identities and capabilities and rely on ad-hoc qualification tasks that fail to respect the workers’ expertise. These practices not only undermine the quality of data but also the ethical standards of AI development. To rectify these issues, we advocate for policy changes to enhance how data annotation tasks are designed and managed and to ensure data workers are treated with the respect they deserve.
\end{abstract}

\section{Introduction}

Much attention has been paid to the creation of datasets, in areas spanning human-centered data science \cite{Aragon_Guha_Kogan_Muller_Neff_2022, mullerHumanCenteredStudyData2019}, to critical reflections on the values shaping dataset construction \cite{dentonGenealogyMachineLearning2021,scheuerman2021datasets}, with an eye to the effects of dataset creation on overall system performance of Artificial Intelligence (AI) and Machine Learning (ML) systems \cite{vasudevan_when_2022}. Efforts in AI ethics, such as the development of transparency artifacts like Datasheets for Datasets \cite{gebruDatasheetsDatasets2021} and similar initiatives (e.g., \cite{CrowdSheets,Healthsheet,Artsheets}), typically require those requesting datasets to provide detailed information about how the datasets were compiled. Crowdworkers who collected or annotated the dataset are best positioned to give these details; however, rather than being seen as domain experts \cite{miceliDataProductionDispositif2022} with valuable perspectives to share \cite{wang2022whose}, crowdworkers are more commonly treated as agents of potential bias \cite{miceliStudyingMachineLearning2022}. This is consistent with the overarching power dynamics in the field of data annotation that work in the requesters' favor \cite{miceli2020between, kapaniaHuntSnarkAnnotator2023}. The oversight of this reality is starkly apparent in the widely acclaimed White House Blueprint for an AI Bill of Rights \cite{OSTP_2022}, which, in focusing exclusively on consumer protection and system accountability, fails to specifically acknowledge the pivotal role of data workers. This omission underscores a significant gap in policy —- the lack of recognition and protection for data workers as integral to the ethical development and operation of AI systems.

The focus on better understanding datasets from workers' perspectives is not misplaced. Data annotators bring a lived experience and perspective to the datasets they work on \cite{denton2021whose}. We struggle with the behavior of systems that generalize off training datasets we don't understand the composition of; such systems introduce significant problems, such as severe discrimination, into AI and ML systems. \cite{barrettSkinDeepInvestigating2023, buolamwiniGenderShadesIntersectional2018, birhane2021multimodal, couldryDataColonialismRethinking2019}. Given the longevity and enduring relevance of these massive datasets \cite{Thylstrup_Hansen_Flyverbom_Amoore_2022, dentonGenealogyMachineLearning2021, Paullada_Raji_Bender_Denton_Hanna_2021}, this issue is a pressing one. Workers have a unique position to observe, identify, and report on the emergence of these trends, e.g., inadequate representation of different subgroups \cite{rolf2021representation}. Historically, though, workers have not been asked to share their perspectives. Instead, they have been actively discouraged from doing so by requesters, as part of the trend towards atomization in digital piecework \cite{grayGhostWorkHow2019, jonesWorkWorkerLabour2021, sambasivanDeskillingDomainExpertise2022a}.

How do we reset the relationship between data workers and requesters? If data workers are treated as domain experts and partners in ethical data collection \cite{jo2020lessons}, their perspectives can enrich efforts to document dataset development and fair use. Such inclusion is not the status quo, however. Further, we do not know how requesters conceptualize workers, their identities, values, commitments, and role in dataset creation for AI and ML systems. By understanding these engagement paradigms, we can better comprehend how the dynamics between requesters and workers not only affect the well-being of the workers but also negatively impact the integrity and utility of the data. 

Our work is thus a large (\textit{n} = 52) interview corpus of requesters active on digital task platforms (e.g., Amazon Mechanical Turk, Clickworker, Appen) who post data tasks as part of the development, testing, and/or fine-tuning of AI and ML systems. As requesters arguably hold the most influence in digital task platform ecosystems---where both platforms and workers rely on them for profits and employment, we turn to them. Prior work, such as the Turkopticon project \cite{iraniTurkopticonInterruptingWorker2013} and other studies of crowd workers, e.g.,  \cite{miceliStudyingMachineLearning2022, hanrahanInvestigatingAmazonMechanical2018, brawleyWorkExperiencesMTurk2016, xiaOurPrivacyNeeds2017}, have documented the experiences of data workers on such platforms and demonstrated platform resistance to change. This suggests a need to focus on the role of other stakeholders (namely, requesters). Subsequently, our study pursues two research questions: \textbf{(RQ1) how do requesters perceive the identity of platform workers, and (RQ2) what are their views on the workers' motivations and work methods}, to enable us to envision strategies for improving worker experience and effectively utilizing the extensive expertise of the workers \cite{Miceli_Yang_AlvaradoGarcia_Posada_Wang_Pohl_Hanna_2022}. 

In shifting the critique of data annotation from those who perform it (workers) to those who request it (requesters), we can see how workers are often rendered invisible in the process\cite{grayGhostWorkHow2019, wang2022whose}. Requesters typically highlight only what they see as `good' data resulting from their tasks, and reference the human workers behind the data only to blame them as `bad actors' or sources of dataset corruption. Recognizing the humanity of these workers is crucial, however. We must embrace it if we want to leverage data annotators' extensive knowledge and dataset understandings.

We report a series of \textit{proxies} that requesters use to identify `good' data and filter out `bad' actors. We define proxies as requirements and pre-task trials designed by requesters to select potential workers and to verify the quality of their work upon completion of the task. The scope and nature of these proxies underscore that dataset production is a process of meticulous curation, a trend increasingly seen as a design practice among requesters. While the proxies range from arbitrary to reasonable, they often overlook the workers' actual lived experiences and their expertise in handling data. As measurement instruments, these proxies fail fairness and bias criteria for `construct reliability' and `construct validity' as adapted by \citet{Jacobs_Wallach_2021}. Consequently, datasets created through processes employing these proxies are potentially unfit for use. Further, as quantitative--and often punitive--measures, they do not foster the respectful relationship with data annotators needed to engage them as skilled dataset contributors.

Our call is not just to reform proxy usage; we urge  pro-social treatment of workers to empower and promote data workers as domain experts within a broader program. As professionals with the most intimate knowledge of the dataset, data workers are best positioned to report on emerging trends or concerns. Our findings demonstrate that the conditions (as set by requesters) needed to foster positive, collaborative engagement between requesters and annotators are \textit{not} present. In our discussion, we suggest sites where positive engagement could be fostered, such as research conferences. To be clear, this is not a policy paper. Instead of proposing new infrastructure or systems, we aim to explicating the status quo of requester--annotation engagement. In doing so, we call upon the research community to collaboratively improve a system that is currently frustrating for both requesters and annotators.

\section{Related Work}

To underscore the significant power disparity between requesters and annotators, we begin by examining digital data work as a form of \textit{invisible} collar labor. Traditional categories of work in the United States include \textit{blue}, \textit{white}, and \textit{pink} collar, each indicating the nature of the work. Blue collar refers to manual-industrial work \cite{Blue-collarworker_2024}, white collar to office and managerial tasks \cite{White-collar-worker_2024}, and pink collar denotes feminized labor, often consisting of underpaid clerical or care roles \cite{howePinkCollarWorkers1978}. \textit{Invisible} collar work represents a new category of labor, characterized by its transient, site-agnostic nature that is typical of many low-status digital roles. This term captures the often unseen and undervalued efforts within the digital workspace. Data annotation, as invisible collar work, is consistent with well-studied power dynamics found in platform-mediated data work, which we detail in the second subsection below. On these platforms, the pseudo-anonymous nature of relationships between workers and requesters results in a focus on quantification as concrete measures of productivity and quality. This contrasts with the comprehensive performance reviews typical in traditional employee-employer relationships. In the final subsection, we connect this trend of quantification to a larger conversation that views data as a product of a design process \cite{feinbergDesignPerspectiveData2017}, rather than an outcome of isolated technical procedures. 

\subsection{Invisible Collar Work and Labor Protections}
Digital task platforms such as Amazon Mechanical Turk (MTurk) often operate under a principle of pseudo-anonymity, a stark contrast to traditional place-based work where supervisors and subordinates share a physical space \cite{xiaOurPrivacyNeeds2017, sannonPrivacyPowerInvisible2019a}. The name 'MTurk' refers to an 18th-century hoax: a  chess-playing machine seemingly automated but secretly operated by a hidden human \cite{Pontin_2007, UntoldHistoryAI2019}. This historical reference illustrates Amazon’s approach to MTurk's workers (or, 'Turkers'), who perform 'human intelligence tasks' (HITs). The design intentionally obscures the humans behind the tasks, creating an illusion of seamless technical efficiency and rendering the workers virtually invisible \cite{vertesiSeamfulSpacesHeterogeneous2014, Tubaro_Casilli_Coville_2020}. This invisibility underscores a critical issue in digital labor, where workers are not merely remote but are systematically hidden from view while contributing essential services in AI development. 

Virtual work lacks the workplace protections commonly found in other employment sectors \cite{jonesWorkWorkerLabour2021}. Crain et al. refer to this as 'invisible labor,' describing it as the often hidden and outsourced online work associated with poor conditions and low pay \cite{crainInvisibleLaborHidden2016}. Unlike \textit{blue} or \textit{white} collar workers, those in 'invisible' collar roles struggle even to be recognized by their de-facto employers \cite{cherry2016virtual}, a critical first step toward receiving workplace protection. Virtual work shares some characteristics with blue collar jobs, e.g. the timed, routine nature of tasks. Other times, it resembles white collar work, being primarily computer-based. However, unlike these traditional categories, \textit{invisible} collar workers lack the protective frameworks: they are not covered by U.S. Occupational Safety and Health Administration (OSHA) regulation that safeguards blue collar workers, nor do they benefit from the compliance mechanism white collar workers have in human resources (HR) departments. Invisible collar work falls through the cracks, consistent with a larger erosion of traditional workplace protections \cite{Bernhardt_Boushey_Dresser_Tilly_2008}. While discussing requester practices, we return to Crain's compendium of the challenges facing invisible collar workers, emphasizing that making these roles more equitable necessitates both legal and policy interventions.

Invisible collar work also mirrors pink collar roles in its physical and psychosocial aspects. \textit{Pink} collar work, often feminized and under-recognized, includes roles like childcare, nursing, or secretarial and administrative labor \cite{howePinkCollarWorkers1978}. Like pink collar workers, invisible collar workers are typically underpaid. However, distance between supervisor and worker is more pronounced, potentially spanning physical locations, time zones, languages, and cultures. This separation also extends to co-workers, complicating both collective organizing and individual actions. Given that computing disciplines are a community largely responsible for generating invisible collar labor, it is our responsibility to ensure proper, pro-social treatment, as advocated by \citet{hawkinsEthicalAmbiguityAI2023} and \citet{rothschildFairProsocialEmployment2022}.

\subsection{Power Dynamics on Data Work Platforms}
\label{platform-dynamics}
As noted in the above subsection, crowdworkers lack traditional methods and recourse of workplace protection. They are directly exposed to the conditions set forth by the platform and requesters alike. Previous efforts such as the Turkopticon \cite{iraniTurkopticonInterruptingWorker2013, iraniStoriesWeTell2016} and Dynamo \cite{salehiWeAreDynamo2015} projects have sought to compel platforms to take on the responsibility to protect and support workers. However, the platforms resisted these changes, ultimately placing the burden back on the workers themselves \cite{iraniStoriesWeTell2016}. Thus, we focus, instead, on the role the requesters play in data work. While platforms should still change for the better, we are more optimistic that requesters can and will do so, particularly with the buy in of policy bodies at the inter-institutional level.

This work examines data work mediated by large platforms (e.g., MTurk), focusing on the relationship between requesters and professionals who complete it, as part of the broader conversation on AI ethics \cite{birhaneForgottenMarginsAI2022}. Data work facilitated by large platforms is characterized by a \textit{triangular} relationship of stakeholders \cite{fieselerUnfairnessDesignPerceived2019}. While labor requesters and performers can sometimes communicate directly, they are often doing so through the platform's mechanisms, sometimes with the platform serving as an intermediary. The flow of information is often one-sided, favoring the requesters. As clients, they command the platform's attention and shape the employment and experience of the worker\cite{hanrahanInvestigatingAmazonMechanical2018, brawleyWorkExperiencesMTurk2016}. Conversely, workers are frequently ignored or actively silenced \cite{miceliDataProductionDispositif2022}. A notable example of this imbalance is Amazon's delayed response to the issue of mass rejections, where requesters reject legitimate work to obtain additional labor at no extra cost, unfairly penalizing workers\cite{ENDHARMMASS2023}. This disparity extends beyond platform treatment to societal perceptions. While data scientists and data analyst receive respect and resources, the data workers who produce the datasets are often undervalued and overlooked \cite{sambasivanDeskillingDomainExpertise2022a, sambasivanEveryoneWantsModel2021b}. 

The treatment of workers on data work platforms is a well-established concern in AI ethics, noted particularly for poor labor conditions and the limited inclusion of their perspectives\cite{diverse-perspectives-can}. Hawkins \& Middelstadt point out that AI researchers are subject to relatively few ethical guidelines, compared to other fields employing crowdworking for purposes of data collection and curation \cite{hawkinsEthicalAmbiguityAI2023}. Barrett et al. question the values and metrics imposed by requesters and their implications in computer vision \cite{barrettSkinDeepInvestigating2023}. While this previous work has focused on the experience of the workers or the platform, we study the practices of dataset requesters to understand how their decisions impact data workers. This insight enables us to highlight the critical role requesters play in shaping working conditions and dataset integrity, thereby informing the development of policy changes aimed at creating just and equitable work environments and ensuring the responsible construction and use of datasets.  

\subsection{Data is Designed}
It is a commonly held idea in critical data studies that data is \textit{designed} \cite{feinbergDesignPerspectiveData2017}. Feinberg's use of \textit{designing data} makes concrete two ideas. First, it formalizes Bowker's notion that data is never \textit{raw} \cite{bowkerMemoryPracticesSciences2005, gitelmanRawDataOxymoron2013}, as any attempt to capture and curate a dataset necessarily imprints existing sociotechnical contexts and choices unto that dataset \cite{Hasselbalch_2021, dignazioDataFeminism2020, batesDataJourneysCapturing2016}. Second, \textit{design} calls attention to the human role in shaping data and datasets, from the collection methods to the analysis software used. In the context of platform-mediated data work, the pseudo-anonymous relationship results in an idea that the data produced is---or can be---`authentic' and `correct,' rather than acknowledging the designed nature of the task and the resulting data. 

Where and how to source `good' data remains a key challenge in the endless quest for Big Data \cite{Mayer-Schönberger_Cukier_2014, boydCRITICALQUESTIONSBIG2012}, as a means of refining, re-tuning, and ultimately improving the performance of ML and AI models. However, previous gold standard datasets have been exposed as sites of error. Among others, significant benchmark datasets, such as \textit{ImageNet} \cite{vasudevan_when_2022, MajorMLDatasetsa} and \textit{80 Million Tiny Images} \cite{Birhane_Prabhu_2021}, remind us of the need to direct attention to how datasets come into being. Having a better sense of where, how, when, and why a dataset is assembled can help ensure that we understand 1) the contents of the dataset itself and 2) help determine reasonable and appropriate generalized use, specifically for the use of AI and ML systems. These datafication moves, as described by Valdivia \& Tazzioli \cite{valdiviaDataficationGenealogiesAlgorithmic2023} often reify existing power imbalances and discriminatory regimes \cite{mittelstadtUnfairnessFairMachine2023, leavyEthicalDataCuration2021}.

Further, the requesters' reliance on quantification to make sense of workers and their work recalls Espeland \& Stevens' \textit{sociology of quantification} \cite{Espeland_Stevens_2008}. They employ computational methods to accept or reject workers, a necessity given the vast scale of datasets required for training AI and ML systems. This approach formalizes the dataset design process. Recognizing this design process is crucial not only for understanding the labor situation of digital pieceworkers \textit{but also} for contextualizing datasets, which aids in determining their appropriate use \cite{gebruDatasheetsDatasets2021, boydDatasheetsDatasetsHelp2021}.

Recognizing the role of all individuals involved in the development of ML and AI systems---from data workers to scientists and engineers---is fundamental to human-centered data science. This approach appreciates the humanistic and subjective dimensions of data science practices \cite{Aragon_Guha_Kogan_Muller_Neff_2022} and considers the implications for how data subjects are represented \cite{chasalowRepresentativenessStatisticsPolitics2021, qadriAIRegimesRepresentation2023}. Understanding the creators of AI systems and their development processes \cite{mullerHowDataScience2019a} is crucial to addressing issues of fairness and bias \cite{gerchickDevilDetailsInterrogating2023}. It is equally important to comprehend how they design and construct the datasets they use \cite{liDimensionsDataLabor2023a, papakyriakopoulosAugmentedDatasheetsSpeech2023}.

Our research integrates these three strands to make a compelling argument: data annotators are relegated to the role of invisible collar laborers; this positioning creates pronounced power imbalances that disproportionately favor requesters. To truly engage data annotators as critical contributors to datasets, it is essential to recognize the influential role of requesters as architects of both the datasets and the conditions under which labor occurs. This understanding is crucial for fostering a collaborative environment that promotes effective information exchange. This is also to our own benefit. When we think of the process of ``flagging'' (or calling attention) to a problem, e.g., in data work, doing so requires interfacing with someone in a decision-making role. Understanding data annotators as invisible collar laborers contributes to complications to the process of flagging: first, because it must be conducted digitally and often indirectly through third party platforms, and, second, because the requester and annotator relationship is marked by a power imbalance, favoring the requester.  

\section{Methods}

To understand the requester--worker relationship, we interviewed 52 participants between November 2022 and January 2023. This was a diverse pool of participants from both academic and industry settings, who acted as requesters for both research projects and commercial products. Our study was approved by our Institutional Review Board (IRB). The first author and 51 students in a graduate-level class on qualitative methods conducted these interviews. While the students in the class produced a total of 87 interviews, we removed interviews that did not meet the study criteria, where the requester was working on an AI or ML task, or if the interview was of low quality. 

The students were well prepared to conduct the interviews after several months of instruction and completing assignments on gathering qualitative data and instruction on human subjects research. Two students participated in each interview to help keep notes and ensure no questions or opportunities for further probing were overlooked. All students had done at least one interview as a class assignment prior to this, and most teams consisted of at least one student with previous human-computer interaction research experience in industry or academia.

The last author taught the course and offered students the opportunity to continue participation in the research project for academic credit; two students elected to do so, with one dropping out and one who is the third author of this paper. Using a real research protocol and tools for analysis is a unique opportunity for students to experience part of an authentic qualitative research practice, which typically takes more than a year to conduct and, thus, can not be completed in a one-semester course. 

Interviews served as the basis of the course's final project, wherein students were tasked with identifying participants (with support from the first and last authors) and conducting a semi-structured interview using the interview protocol created by the authors. The objective of the research and interview protocol was thoroughly explained to the students before they began work and all students were part of the IRB's approved human subject protocol. They were encouraged to speak to a requester who had posted a task for data annotation, with each student expected to find and lead one interview.

\subsection{Participant Selection}
The criteria for participant selection was to include industry and academic professionals who have used a crowdsourcing platform (e.g., Amazon MTurk) to source data or data work for use in a large, data-intensive system (mainly for AI and ML). Potential participants were identified by the students through a range of sources, e.g., forums for particular digital task platforms and students’ professional networks, along with those of the course teaching staff. Students recruited participants through word of mouth, email and social media posts. While the sample size of 52 was large for an interview study, we recognize that patterns in recruitment, geography, and shared professional networks were not representative of all types of requestors. We do not intend for our participants' experiences to be seen as representing \textit{all} requesters, rather they serve as a broad sample of requesters from different organizations. Participants were not compensated in any way for their participation in the study, due to Georgia Institute of Technology regulations.

\subsection{Interview Structure}
Semi-structured interviews \cite{edwardsWhatQualitativeInterviewing2013}, following the interview protocol, were conducted with all the participants. This was followed by a 10 minute think-aloud session where participants walked the interviewer through a task on a platform they had used for crowdsourced data collection or annotation. All interviews were conducted via Zoom. On average interviews lasted about 45 minutes. At the end of the interview, participants were tasked to fill out a demographic survey (age, gender, employment status, and educational background). They also filled a separate form to express their consent (if granted) to be contacted for follow-up interview. Interviews were then transcribed and anonymized. 

Interview questions focused on the participants' experience of crowdsourcing and perception of the workers who completed their task(s). Participants were also prompted to share more about certain nuances like payment calibration and metrics to qualify workers for a task, as part of the think-aloud session.

\subsection{Data Analysis}
The first author developed a preliminary codebook after reviewing all interviews and performing open coding \cite{flickDoingGroundedTheory2018}, in consultation with the last author (course instructor), who had listened to each student group present their findings from their interviews as part of the final project. The whole research team then read a selection of interviews and updated the codebook, before the first and second authors re-coded the entire interview corpus with the new codebook. Throughout the re-coding process, incremental changes to the codebook were made in consultation with the larger research team through weekly meetings over about three months. After the re-coding was performed, the research team discussed emerging themes and refined the open codes via axial coding, based on the grounded theory approach \cite{straussGroundedTheoryPractice1997}.

\subsection{Participant Backgrounds}
We highlight four aspects of participants' backgrounds, as known to us, to help contextualize their experiences:

\textit{Work context}: We roughly delineate participants into ``research’’ (38) and ``commercial’’ (14) requesters. This distinction is based on the purpose of the tasks the requester makes; ``commercial'' tasks involve fueling or refining a commercial product (e.g., a voice technology startup). Fundamental knowledge or insight is grouped into ``research’’ requests, e.g., at a university lab, to test the naturalness of generative text systems. From this background factor, we can make sense of where interventions might be most useful (e.g., research conference vs corporate guidelines). 

\textit{Learning how to `request'}: We are concerned with how requesters come to understand a platform and design tasks for it. One site of intervention might be in the materials requesters use to learn how to create a task on a platform. Several participants used multiple resources: co-workers or collaborators were a frequent source of instruction (17), while another 15 participants turned to search engines and discovered articles, blog posts, and GitHub repositories with suggestions. The same number (15) referenced platform-published documentation. Another nine participants learned from YouTube tutorials, with eight receiving instruction from their academic advisors. A fair number (7) consulted no explicit instruction materials and instead learned from self-exploration of the platform. A much smaller number (3) received explicit support from the company operating the platform.

\textit{Estimating task timing}: Task timing provides some preliminary insight into how participants made sense of the workers, though not all participants described their model for estimating task timing. For example, how requesters compare their own reading speed to that of potential workers. The majority of participants (12) did a rough, self-determined estimate of how long it would take workers, while nine asked their fellow lab members to try the task and average their time taken. A few (3) conducted a formal pilot test with workers to get an average completion time.

\textit{Calibrating payment}: The most common model (11) for determining payment was paying their local minimum wage (often \$12/hour) based on how long they thought the task took to complete, with an additional two paying ``slightly more'' than minimum wage. Some participants determined how much to pay workers based on direct instruction from a superior (advisor or boss) (2) or as derived from total budget divided by number of tasks (3). Others (5) differentiated based on what they saw as the types of questions or tasks, to which they had a pre-assigned mental model for appropriate payment. Several participants (11) assigned a random amount of payment (e.g., 50 cents) they felt was appropriate, or assigned a random number. A small number (2) followed what the platform suggested (amount varied by platform).  

Participants requested tasks in three main categories (interviewees may have multiple kinds of task requests). \underline{Data annotation and classification} (30 total): Including text annotation (e.g., sentiment annotation, or rating violent intention in social media posts) and image labeling (e.g., drawing bounding boxes, or describing image contents). Generally refining or labeling an existing dataset for use in model training. \underline{Data collection} (12 total): Used to procure a dataset for model training, including tasks like requesting YouTube viewing history and collecting speech samples and text generation for NLP projects. \underline{Tool or system feedback} (19 total): User testing or reviews of AI or ML systems. Sometimes of a non-standalone algorithm, such as rating accuracy of image classification system that will be incorporated into a larger product. Other examples include rating or describing experience with a chatbot, Explainable AI system, or the naturalness of a generative NLP system.

Participants used various digital task platforms and several participants active on multiple. Platforms used include Amazon MTurk (46), CrowdFlower \& FigureEight \& Appen\footnote{CrowdFlower was renamed FigureEight, which was then acquired by Appen in the span of a few years; we report these in a group since some participants referred to the company by its name at the time of requesting and others by its current name.} (8), Prolific (5), Microworkers (1), Upwork (1), Fiverr (1), Dial Crowd (1), Labeler (1), Tokola (1), iMerit (1).

\section{Findings}

We present our findings in three sections: the first covers how requesters perceive the workers they hire on digital task platforms. Second are the proxies requesters use as worker selection criteria. Third, the curated nature of datasets emerging from the dyadic requester-worker interactions, focusing on how proxies systematically exclude workers' perspectives and affect data quality.

\subsection{Requesters' Perceptions of Worker Identity}
We saw significant variation in the vocabulary requesters used to describe platform workers. The workers employed by our participants were referred to as ``users,'' ``participants,'' as well as ``workers.'' Interestingly, individual requesters often employed several of these terms interchangeably during their interviews while describing the same group of workers. The term ``employee'' was notably used just once (among the 52 interviews), by P42, to draw a strict distinction between platform workers and the full-time annotators (referred to as ``employees'') at P42's company.

\subsubsection{Who requesters think workers are}
Two prevalent paradigms emerged for hiring or determining worker eligibility for tasks on digital task platforms. On non-anonymous platforms like Upwork, where workers have a clear identity or personal brand visible to requesters, selection hinges on the worker's portfolio and profile. In contrast, on pseudo-anonymous platforms, such as MTurk, requesters know almost nothing about the worker. However, they can either apply platform sponsored filters, or create their own screening tasks to target specific groups, like college students. This subsection focuses on the latter model, specifically the requester-led screening processes typical of most pseudo-anonymous digital task platforms such as MTurk. 

Workers on digital task platforms can theoretically be anyone with an internet connection and a suitable computer, subject only to platform-specific requirements regarding country of residence, age, and language skills. While the exact demographic and experiential composition of these workers has been estimated in various studies, we found that requesters hold only a vague understanding of the potential worker pool. This observation aligns with what Kapania et al. observed \cite{kapaniaHuntSnarkAnnotator2023}. Many requesters perceived the worker base of digital task platforms as ``the general public''. For instance, P31 explained that they imposed no specific restrictions on which workers could complete on their tasks, aiming for \textit{``the general workers.''} Similarly, P36 admitted to having \textit{``no idea''} who was completing their tasks, describing the workers as \textit{``random''}. P6 echoed this sentiment, deeming worker background insignificant to their task and their relationship with the workers as \textit{``strangers on the platform.''} 

A minority of requesters had specific (albeit often anecdotal) insights into their worker population. For example, P1 communicated regularly with a few Amazon MTurk workers who were on government disability benefits and supplemented those by working from home on MTurk.From these interactions, P1 formed a general impression that many of their workers had similar circumstances. However, they recognized that their view might be skewed, noting, \textit{'[I] mostly imagined those people even though I know they're not the majority...they're just the ones who communicate more.'}"

Most requesters, however, stuck to the idea that workers on these platforms represented the general public. These requesters then implemented various measures to secure ``good data.'' Participants were frequently concerned with the quality of the data (or task performance), as it contributes to the larger outcome dataset and subsequent model development. To get ``good'' data, requesters took strenuous efforts to protect against the biggest perceived threat: ``bad actors''.

\subsubsection{Good data vs bad actors}
Though participants had only vague notions of who was completing their tasks, they expressed significant concerns about these ``bad actors,'' or workers who submitted poor or fraudulent responses just to receive payments. Notably, while the data quality produced by a worker might be deemed ``good,'' individual workers were never described in any of the interviews as ``good''. This implies that requesters, when content with the data quality, did not question the identity of the worker. For example, P5 believed that offering a minimum of \$20 per hour would ensure  \textit{``good annotations''}. This reflects a common belief in freelancing that higher pay attracts better candidates. However, P5's focus was solely on the data quality, not the characteristics of the person producing it.  

In contrast, participants dissatisfied with the data they received cited ``bad workers'' whom they saw as performing the tasks poorly or maliciously. In these cases, requesters attributed the data quality to the \textit{worker's} shortcomings. For instance, P9 spoke about implementing quality checks to avoid workers who \textit{``usually cheat in these platforms.''} P39 explicitly described methods for \textit{``filtering out bad actors,''} implying that workers were responsible for undesirable data. Similarly, P13, who usually accepted most all submissions, stated that their only reason for rejection was when \textit{``someone deliberately tries to poison the data.''} P13 continued describing their methods for manually reviewing task submissions to determine which workers \textit{``are doing bad.''} 


Many academic requesters, who posted components of academic studies as tasks, perceived workers primarily motivated by financial compensation as a threat to research integrity. These academics (often implicitly) expected workers to have a sincere interest in contributing to the development of new knowledge. Even though P44 recognized that \textit{``Amazon Turkers is a group of people that does this for money,''} they were frustrated by workers or research participants, who they felt were \textit{``just clicking, you know, like randomly or very quickly, to get you [the requester] the [completed] survey to get money.''} This is only one telling example of the conflict between the academic expectation of earnest participation and the practical motivation these platform workers.

The \textit{good data vs bad actors} dichotomy reflects a broader dynamic in the relationship between requesters and workers. When task submissions, as contributions to a dataset, meet expectations, requesters typically don’t question the workers' identities, expertise, or experience. Yet, if the data is unsatisfactory, the focus shifts to the workers as individuals. Suddenly, the requester holds the worker (as a \textit{bad actor}) responsible for the subpar outcomes. This shift underscores how workers' human identity and experience are overlooked when things go well, but scrutinized when issues arise.

\subsection{Proxies for Selecting Works and Assessing Quality}

To discern workers who would produce high-quality data from potential bad actors, our interviewees developed a number of \textit{proxies}, or micro-tests to approximate whether a worker holds or delivers a certain quality (see all proxies in a summary table in Appendix \ref{tab:proxies}). There are two types of proxies: 1) those determining a worker's identity and suitability for a particular task, and 2) those evaluating their ability to produce high-quality data. Notably, many requesters used multiple proxies in tandem, filtering out elements of what they believed to be the general (implied US) population. For example, P11, who wanted to recruit members of the general public for their task, expressed a desire to avoid biases such as English fluency in their task outcomes. However, P11 used a number of proxies, including filtering potential workers based on their prior task acceptance rate. 

\subsubsection{Proxying worker identity}
The following is a synopsis of the proxies used to establish the workers' identity on these quasi-anonymous data work platforms. 

\textit{English fluency}: Interviewees (all U.S.-based) frequently emphasized the importance of potential workers' English fluency. P30, requesting data collection for NLP tasks, prioritized hiring \textit{``English native speakers''} and employed platform filters to exclude workers accordingly. In contrast, P19 described their strategy for ensuring English fluency: selecting workers from locations where English is a primary language. Though they admitted to not finding a reliable method on MTurk \textit{``to ensure that that requirement is met.''} 

\textit{Age}: Due to legal requirements, many interviewees stipulated that workers must be at least 18 years old. In academic tasks, requesters typically asked workers to simply affirm they were were 18 or older, a standard practice for U.S.-based human subjects research. Conversely, in commercial or industrial tasks, interviewees more often relied on platforms' official filters and purview for age verification. 

\textit{Location and time zone}: Location-based proxies were popular among requesters doing geographically-specific projects. E.g., P16 wanted workers familiar with U.S. news and events and limited their selection to U.S. residents only. Similarly, P18, who was interested in having data annotated with specific English dialects (UK, Australia, Canada, Ireland, New Zealand), selected for workers from these countries using platform filters. However, P18 also knew workers could use VPNs to appear as if they were in these countries. P37, who also sought workers from the UK, Australia, and the U.S., posted tasks during regular working hours in these regions, instead of the requester's own local timezone. P37 also informally reviewed long-form text submissions believing \textit{``from the way they [the workers] speak,''} they could identify the worker's \textit{``ethnic background''} and whether it aligned with their targeted regions or not.

\subsubsection{Proxying aptitude}
Requesters employed proxies to ascertain how well a given worker would complete---or did complete---a given task. The proxies are split into pre and post hoc, or those applied to test worker aptitude based on their platform-specific track record or those used to test how well a worker performed on the requester's task. 

\underline{Pre} hoc proxies:

\textit{Approval rating}: Requesters commonly used approval ratings, which reflect a worker's aggregate ratings from tasks completed for other requesters, as a filtering criterion. However, what constituted an acceptable prior approval rating varied between requesters: P47 considered a rating above 95\% as \textit{``good''}, while P10 set their minimum at 98\%, and for P11, the threshold was even higher at 99\%. 

\textit{Number of tasks completed}: Requesters often gauged worker's reputation by their previous task history, i.e., the number of tasks they completed. Similar to approval rating, the threshold for desirable task history varied greatly. For P21 that number was 1,000 prior tasks, where P48 set a benchmark at 100, and P10 at merely 50.

\underline{Post} hoc proxies:

\textit{Keyboard interaction}: Requesters calculated how much a worker engaged with the keyboard---and the speed of keyboard interactions---to ascertain how sincerely workers completed their tasks. P16 wrote a tool to check if workers copied and pasted into free-text boxes, in order to ensure that data collected \textit{``was not useless.''} P12 assumed that a fast repetition of clicks meant the worker was a bot, and so employed a keylogger to make sure workers were genuine with their replies. P12's concern was workers just \textit{``trying to get their two cents or whatever.''} P39 performed an \textit{``entropy analysis''} of the amount of time spent on the task, premised on the idea that if the worker spent far less time than the team determined appropriate in trials, the worker's submission would be rejected.

\textit{Answer pattern}: Requesters saw the visual pattern of answers or choices (on sequential multiple choice and Likert scale questions) that workers selected as being a signal of genuine effort. For example, if a worker always selected Option B in a multiple choice question, that was a red flag for P12. Similarly, P3 would manually review all task submissions, checking if participants choose only the first option, or another pattern.

\textit{Attention check}: Requesters used assorted wording (e.g., \textit{``quality checks'')} [P9], \textit{``dummy questions''} [P52]) for what is commonly known as an \textit{attention check}. These questions verify is not a bot and is focused on the task at hand. Requesters commonly complained that grading attention check results was time-consuming, and perhaps \textit{``the most time-consuming activity''} of posting a task (P9). P34 explained that they began including attention checks because they \textit{``didn't trust them [the workers] anymore.''} CAPTCHAs are commonly used to filter out bots, while reading comprehension questions are often used to make sure workers are carefully reading directions. For example, P1 tried to write what they felt were fair attention checks, or ones that are \textit{``really, really obvious where you can only get them wrong if you're literally not reading...it's a multiple choice question that says do not choose the other option, the other options says you have to choose this option.''} Perhaps the best summation of the nature of attention checks is the experience of P21, who used a series of attention check methods and, in turn, received feedback from workers in which they called P21 an \textit{``evil genius''} due to the difficulty of properly completing the task.

\textit{Gut reaction}: Several requesters relied on a \textit{gut reaction}, or instinctual feeling of whether or not task submissions were correctly done. P36 had a two-part strategy for verifying the quality of worker submissions, the second step being \textit{``more or less ad hoc...[it] was subjective...because we would look at the annotations and be like, `okay, yeah, this person has been producing consistently good annotations, so he should continue with it.'''} P50, meanwhile, described going through submissions for a generative task manually and deciding \textit{``these ones are good, these ones are bad,''} a process they ultimately found \textit{``annoying.''} 

\textit{Coherence as trust}: Comparing a worker's answer or selection for a subset of task activities was a common way to determine how correct a submission was. Two methods were used to do this. In \underline{majority consensus}, requesters pick the most commonly submitted label or annotation and check all submissions against that correct answer, rejecting submissions that don't comply. P2 set a threshold for agreement and once that agreement was met, accepted that label as the correct one. P16 similarly employed agreement scores between workers to determine the acceptable annotation. The other method was \underline{gold standard adherence}. P5, for example, checks the agreement score of different annotators against their (requester developed) answers and if the score is high enough, accepts the submission. Similarly, P50 used a small pool of self-labeled data to test worker submissions against.

Another indicator of potential worker submission quality used by requesters operating on MTurk was the Amazon-designed \textit{Master Turker} distinction. Of the qualification, Amazon vaguely states that Master Turkers have completed thousands of tasks on the platform and have done so with a high level of performance.\footnote{\url{https://aws.amazon.com/blogs/aws/amazon-mechanical-turk-master-workers/}} Notably, hiring only Master Turkers as a requester on MTurk carries an additional platform-imposed surcharge. Requesters in our corpus had mixed feelings about Master Turkers and what the qualification meant about workers holding it. Some felt it was an assurance of reputable work; P51 posted tasks only for Master Turkers, feeling their work quality was far superior to those of non-Masters. P20 felt the surcharge for Master Turkers was worth it, since \textit{``we ran a cost analysis and realized that it was actually cheaper to pay significantly more for Master Turkers because the results that they gave us back were just so much better.''}

Others, however, felt the designation to be relatively meaningless. P37, for example, stopped using Master Turkers after reading on a forum that the same level of worker could be selected for using a mix of other filters, including prior approval rating. P18 felt that using only Master Turkers was contrary to their goal of capturing a broad cross-section of workers; they wanted a wider annotator pool that what they would find in an academic research environment. P18 felt Master Turkers represented an elite within the larger MTurk worker population. Similarly, P5 was not sure the Master Turker designation was meaningful---they were also concerned with workers being unsure of how to obtain the designation, which they saw as unfair.

\subsection{Curated Datasets}

Reflecting on the larger patterns of data curation as observed in our findings, we identify a composite insight into the behaviors of requesters posting ML and AI tasks on semi-anonymous digital task platforms. These requesters attempt to ascertain the following:
\begin{itemize}
\item task-relevant information about a worker's identity,
\item a promise of quality based on past performance,
\item confirmation of the quality, authenticity, or sincerity of worker submissions, varying by task type.
\end{itemize}

Requesters typically view workers through one of three lenses. Firstly, some requesters are indifferent to workers' identities, focusing solely on task completion, which raises concerns about the datasets' ability to represent diverse populations accurately. Secondly, others assume their workers represent the \textit{general public} or specific subgroups without concrete evidence, relying instead on outdated studies. Thirdly, some requesters perceive workers as voluntary study participants rather than paid employees, highlighting a fundamental misunderstanding that impacts mutual satisfaction.

The use of proxies poses challenges for both requesters and workers. Proxies are intended to approximate worker identity and labor quality, but developing effective proxies is problematic due to the lack of standardized methods. For instance, some requesters resort to copying methods from academic papers (e.g., P28) or seek advice from online forums (e.g., P37) to establish these measures. Despite these efforts, workers often find themselves needing to re-qualify for tasks they are already skilled at, wasting time and effort. Moreover, proxies may inaccurately measure the real value that requesters seek. For example, the preference for 'native' English speakers over fluent speakers raises questions: Why would an immigrant in a majority English-speaking country not be fluent enough for the task? This overlooks the richness of varied English vernaculars, which could be valuable in large-scale ML or AI applications.

Workers face additional burdens from proxies, having to prove their competencies repeatedly, which may not reflect their actual domain expertise. Proxies may also be a poor test of actual domain experience and knowledge. Consider the minimum time limits many requesters use to track the sincerity of worker effort. If a worker spends much of their day (given that many digital task platform workers do so as their main income source \cite{hitlin_research_2016}), developing a particular skill e.g., drawing bounding boxes for image labeling, they are highly likely to complete it far quicker than a requester who has perhaps drawn only a few for practice. Workers must also spend time qualifying for tasks through pre-screeners---while some requesters still compensated workers for these pre-task activities, many more did not.

These proxies also force workers to engage in sub-standard work should their ratings fall due to simple misunderstandings, or even malicious requester actions. The Turkopticon ``End the harm of mass rejections'' campaign, for example, highlights how requesters can take advantage of workers, basically by collecting their labor and failing to pay them for it, forcing workers to pick up ultra-low paying tasks in order to rebuild their ratings \cite{ENDHARMMASS2023}.

Finally, the implementation of proxies can lead to the exclusion of valuable data, as requesters curate what they consider undesirable submissions. This selective process contradicts the goal of capturing a broad spectrum of inputs from the general public, limiting the datasets' authenticity and usefulness.

\section{Discussion}

In this paper, we've highlighted significant issues in how digital task platform workers are evaluated. Workers face vague, scale-driven proxies that not only disadvantage them but also compromise dataset integrity. These datasets, often skewed by the underrepresentation of minority voices, form the basis of AI and ML systems, posing risks to both the general public and developers. Moreover, the lack of standardization in these practices makes them unreliable, reinforcing 'the invisible collar' around data annotators. In subsequent sections, we propose strategies to safeguard worker rights and assess the implications of sourcing datasets, emphasizing the use of existing checks-and-balance systems for efficient change implementation.

\subsection{The Pitfalls of Proxies}

Returning to the impact of quantification in platform-mediated data work, as discussed by \citet{Espeland_Stevens_2008}, proxies emerge as a prime example. At a fundamental level, proxies present a concerning practice: as validation implements, they are of questionable \textit{construct reliability} and \textit{construct validity}. 

Drawing on Jacobs \& Wallach's exploration of fairness and bias, \textit{construct reliability} is akin to replicability \cite{Jacobs_Wallach_2021}. Applied to proxies, there is little replicability among requesters, as each implements a conceptual proxy differently. Within our corpus, there were no universal practices for any given proxy. Instead, requesters often rely on practical wisdom from professional or academic sources to create their own interpretations, leading to significant variance in how they establish whether or not a worker has a quality, such as perceived English fluency. For example, some requesters rely on a platforms own characterization (or label) of a worker's language proficiency, while others allow only workers in majority English-speaking countries to complete their tasks.

Proxies not only fail \textit{construct reliability} but also lack robust \textit{construct validity}. According to Jacobs \& Wallach, \textit{construct validity} entails ensuring that measurements are accurately grounded in the intended construct and encompass all its relevant aspects \cite{Jacobs_Wallach_2021}. The proxies requesters used do not meet validity in this regard, as there is no consensus or evidence confirming these measures accurately and comprehensively represent worker's identity or the quality of task completion. These proxies seldom offer a fair assessment of labor. Continuing with the example of English-language skill, some requester priorize \textit{native} speakers over \textit{fluent} speakers, failing to recognize that fluency and native-speaking are related yet distinct concepts. Meanwhile, others strive for a general consensus, which can inadvertently suppress diverse perspectives. This approach can lead to the silencing and even penalization of workers who express minority opinions or unique lived experiences, as noted by Kapania et al. \cite{kapaniaHuntSnarkAnnotator2023}. Further, domain experience is paradoxically undervalued, as seen when workers who complete tasks rapidly are excluded from the final dataset. This raises significant concerns about the fairness and efficacy of proxies in accurately assessing and valuing the contributions of platform workers.

Proxies demonstrate several problems with requester understanding of workers and the work workers perform. As a tool, proxies are meant to check whether work will be (or has been) completed correctly; however, the proxy exists because the requester does not know how the work is done. Proxies also lack both construct reliability and validity. Proxies also impose unfair expectations on workers. Requesters frequently use attention checks and review visual answer patterns, expecting a level of focus from workers that is often unattainable or simply inaccurate. This raises an important question: Why is there an expectation for digital task platform workers, many of whom rely on these platforms as their primary income source, to maintain intense concentration for their entire workday or on each micro-task? It’s unlikely that these requesters themselves maintain such hyperfocus throughout their own multi-hour workdays. This disparity highlights a potential double standard: Why is it deemed fair to expect digital task platform workers to exhibit a level of focus that is not expected in other (lowly compensated) professional contexts? This contrast demonstrates once again the power imbalance between data work annotators and requesters \cite{miceli2020between, wang2022whose}.

\subsection{Labor Protections for Invisible Collar Workers}

Requesters on digital platforms often view workers as primarily driven by financial gain. This misperception limits workers' incentives to exceed minimum quality standards, contrasting sharply with the paradigm in citizen science where volunteers are driven by personal fulfillment or a commitment to a broader mission, often resulting in higher quality data contributions \cite{maund2020motivates}. However, digital platform workers typically face limited job security and scant prospects for advancement, leading to a diminished sense of ownership and investment in their tasks. 

The extensive use of proxies highlights the urgent need to reevaluate the requester-worker relationship, especially acknowledging the absence of labor and workplace protections for data workers on these platforms. As part of the 'invisible collar' class, these workers, despite often earning regular incomes, are not recognized as employees by platforms or requesters. This lack of recognition exacerbates the invisibility of their labor, identity, and critically, their rights and protections. 

Recognizing digital task platform workers as legitimate employees is crucial for extending labor protections to them. While protections for white collar workers involve compliance mechanisms (e.g., HR) and blue collar protections include safety regulations and the ability to unionize, these are not perfect. For instance, union activity can face corporate resistance, and HR departments may prioritize legal compliance over employee well-being \cite{Sainato_2023}. However, any protection is better than the current situation for invisible collar workers, who lack such safeguards.

Furthermore, the White House Blueprint for an AI Bill of Rights \cite{OSTP_2022} fails to address the role of data workers, despite its focus on improving AI and ML system oversight. This oversight is particularly concerning given the significant risks associated with such work, highlighted by the severe impacts reported by data workers in Kenya \cite{roweItDestroyedMe2023} and ongoing legal challenges involving major tech companies \cite{njanjaMetaBarredKenyan2023, MetaAccusedHuman}. As consumer protections in AI increase, such as those seen in the EU AI Act, it is imperative to extend similar safeguards to data workers, recognizing their critical role in the AI ecosystem.

\subsection{Rethinking Dataset Sourcing Policies at Scale}

Following the discussion on the need for better protections for data workers, it's important to recognize that requesters from different sectors face varied pressures that influence how they engage with these workers. Academics are typically governed by the standards of publication venues, Institutional review board (IRB) requirements, and funding body regulations, such as those from the U.S. National Science Foundation, which focus on research integrity and compliance. In contrast, industry researchers are influenced by corporate policies, which prioritize business objectives and operational efficiencies. These sector-based differences underscore the necessity for comprehensive policy interventions that span both academic and industry domains to ensure consistent and fair treatment of data workers across the board. To bridge these gaps and create a unified standard, we advocate for the implementation of industry-wide policies that cover all institutions involved in data work. The following measures are proposed ways to safeguard data workers and enhance the fairness and integrity of digital platforms but should be subject to further study:

\begin{itemize}
    \item \textbf{Standardization of Proxies:} Develop industry-wide standards for the use of proxies in platform-mediated data work. These standards should focus on improving construct reliability and validity, ensuring that proxies are replicable across different requesters and accurately measure what they claim to. Guidelines should be established for developing, testing, and revising proxies to ensure they are fair and effective.
    \item \textbf{Transparency Requirements:} Implement policies that require requesters to disclose the criteria and rationale behind their proxy measures. This would enable workers contest unfair or inaccurate proxies.
     \item \textbf{Worker Involvement in Proxy Design:} Encourage or require the involvement of worker representatives in the design and review of proxies. This involvement can ensure that the proxies reflect the actual expertise and circumstances of the workers.
     \item \textbf{Regular Audits and Reviews:} Introduce regular audits of proxy use by an independent body to ensure compliance with fairness and reliability standards. 
     \item \textbf{Protection Against Proxy Abuse:} Establish protective measures for workers against potential abuses of proxy measures, such as unwarranted rejections or unfair task allocation. This could include mechanisms for workers to appeal decisions made based on proxy assessments.
     \item \textbf{Compensation for Qualification Tasks:} Mandate compensation for workers’ time spent on qualification tests or pre-task screenings. This policy would acknowledge the time and effort workers invest in accessing work.
     \item \textbf{Reduction of Over-reliance on Proxies:} Encourage requesters to supplement proxy measures with other forms of evaluation where possible, such as direct feedback, trial periods, or continuous performance evaluations.
\end{itemize}

\section{Limitations}

Our work is centered on the perspectives \& experiences of requesters working in the U.S. This study population was  intentional, as it is reflective of American academic and industry training on how to serve as a requester on platforms like MTurk, but it is not necessarily indicative of practices globally. Consistent with traditional limitations of interview-based work, we report on interviewees' own experiences and perspectives in their own words, which may not be a perfect representation of how they actually behave in situ.

\section{Conclusion}

Traditionally, data scientists and system developers prioritize the technical aspects of dataset construction. However, our interviews with 52 digital task platform requesters highlight the importance of understanding requesters' perspectives on worker identity, skill, and qualification. Despite constant proxy-based evaluations of their qualifications, productivity, and work quality, we found that workers remain largely invisible unless the requester is dissatisfied with their task performance. These proxies compromise the integrity of worker evaluations and data quality, as they lack reproducibility, objectivity, and standardization. Such practices obscure workers' experiences and perspectives, contributing to broader patterns of worker erosion in dataset curation. Understanding these issues is crucial to addressing the oversight of worker status. Consequently, we advocate for the revision of dataset sourcing policies in both academic and industrial settings, focusing on improving the treatment of data workers and advancing AI ethics initiatives.

\section{Research Ethics and Social Impact}

\subsection{Ethical Considerations Statement}
As this paper details work performed with qualitative methods (namely interviews), our research ethics concerns are primarily related to protecting interviewee privacy. Our entire protocol and data management plan was approved by [institution name]'s IRB, including a waiver of informed consent that explained the study's purpose and how the data collected would be used and stored. Participants' names were not collected by the research team, unless they consented to being potentially contacted for a follow up interview, in which case their name and contact information was stored in encrypted storage, and has already been deleted (following completion of data analysis). Other precautions included following standard procedure for protecting the identity of participants, including storing all direct transcripts in encrypted storage to which only the research team had access, with all researchers having completed [institution name]-sponsored IRB approved human subjects training. Further, in the interviews themselves, interviewees were asked not to share their organizational affiliations (besides describing them as academic or industrial), to avoid creating repercussions for any participant who spoke contrary to their organization's official policies or without organizational approval.

\subsection{Researcher Positionality Statement}
Additionally, our positionality is that of academic and industry researchers who have worked as both data work platform requesters and workers. Our disciplinary backgrounds are diverse, ranging from learning sciences and critical data studies to health informatics and wellness and responsible AI. Several authors on this paper have worked on platforms for data annotation as crowd workers, including the first author who has completed more than 200 tasks on several major platforms, including AMT.

\subsection{Adverse Impact}
The main concern, with regards to adverse impacts, that we can imagine stemming from this paper is professional repercussions aimed at participants who were interviewed as part of this work. We have made strenuous effort to protect identities of all participants to avoid this. 

Another concern is that this work could be misconstrued to suggest that data work should not happen on digital platforms. This is the not the case---our argument is that requesters need to recognize their responsibility in acting as employers to platform data workers and, in many cases, reevaluate the way they employ those workers.

\section{Acknowledgments}
We thank our anonymous reviewers for their feedback on this work. Lauren Klein gave helpful feedback on an early iteration of this work. This work is supported by NSF grant \#1951818 DataWorks: Building Smart Community Capacity and a Google collaboration gift, ``Examining the data practices of human-in-the-loop ML development''. Any opinions, findings, conclusions, or recommendations expressed in this material are those of the authors and do not necessarily reflect the views of the National Science Foundation or other supporters.

\bibliography{aaai24}

\begin{thebibliography}{83}
\providecommand{\natexlab}[1]{#1}

\bibitem[{Aragon et~al.(2022)Aragon, Guha, Kogan, Muller, and Neff}]{Aragon_Guha_Kogan_Muller_Neff_2022}
Aragon, C.; Guha, S.; Kogan, M.; Muller, M.; and Neff, G. 2022.
\newblock \emph{Human-centered data science: an introduction}.
\newblock Cambridge, Massachusetts London, England: The MIT Press.
\newblock ISBN 978-0-262-54321-7.

\bibitem[{Barrett, Chen, and Zhang(2023)}]{barrettSkinDeepInvestigating2023}
Barrett, T.; Chen, Q.; and Zhang, A. 2023.
\newblock Skin {Deep}: {Investigating} {Subjectivity} in {Skin} {Tone} {Annotations} for {Computer} {Vision} {Benchmark} {Datasets}.
\newblock In \emph{Proceedings of the 2023 {ACM} {Conference} on {Fairness}, {Accountability}, and {Transparency}}, {FAccT} '23, 1757--1771. New York, NY, USA: Association for Computing Machinery.
\newblock ISBN 9798400701924.

\bibitem[{Bates, Lin, and Goodale(2016)}]{batesDataJourneysCapturing2016}
Bates, J.; Lin, Y.-W.; and Goodale, P. 2016.
\newblock Data journeys: {Capturing} the socio-material constitution of data objects and flows.
\newblock \emph{Big Data \& Society}, 3(2): 2053951716654502.
\newblock Publisher: SAGE Publications Ltd.

\bibitem[{Bernhardt et~al.(2008)Bernhardt, Boushey, Dresser, and Tilly}]{Bernhardt_Boushey_Dresser_Tilly_2008}
Bernhardt, A.; Boushey, H.; Dresser, L.; and Tilly, C. 2008.
\newblock An Overview of the Gloves-Off Economy: Workplace Standards at the Bottom of America’s Labor Market.
\newblock \emph{Center for Social Policy Publications}.

\bibitem[{Birhane and Prabhu(2021)}]{Birhane_Prabhu_2021}
Birhane, A.; and Prabhu, V.~U. 2021.
\newblock Large image datasets: A pyrrhic win for computer vision?
\newblock In \emph{2021 IEEE Winter Conference on Applications of Computer Vision (WACV)}, 1536–1546.

\bibitem[{Birhane, Prabhu, and Kahembwe(2021)}]{birhane2021multimodal}
Birhane, A.; Prabhu, V.~U.; and Kahembwe, E. 2021.
\newblock Multimodal datasets: misogyny, pornography, and malignant stereotypes.
\newblock arXiv:2110.01963.

\bibitem[{Birhane et~al.(2022)Birhane, Ruane, Laurent, S.~Brown, Flowers, Ventresque, and L.~Dancy}]{birhaneForgottenMarginsAI2022}
Birhane, A.; Ruane, E.; Laurent, T.; S.~Brown, M.; Flowers, J.; Ventresque, A.; and L.~Dancy, C. 2022.
\newblock The Forgotten Margins of AI Ethics.
\newblock In \emph{Proceedings of the 2022 ACM Conference on Fairness, Accountability, and Transparency}, FAccT ’22, 948–958. New York, NY, USA: Association for Computing Machinery.
\newblock ISBN 978-1-4503-9352-2.

\bibitem[{Bowker(2005)}]{bowkerMemoryPracticesSciences2005}
Bowker, G.~C. 2005.
\newblock \emph{Memory practices in the sciences}.
\newblock Inside technology. Cambridge, Mass: MIT Press.
\newblock ISBN 978-0-262-02589-8.
\newblock OCLC: ocm60776866.

\bibitem[{boyd and Crawford(2012)}]{boydCRITICALQUESTIONSBIG2012}
boyd, d.; and Crawford, K. 2012.
\newblock {CRITICAL} {QUESTIONS} {FOR} {BIG} {DATA}: {Provocations} for a cultural, technological, and scholarly phenomenon.
\newblock \emph{Information, Communication \& Society}, 15(5): 662--679.

\bibitem[{Boyd(2021)}]{boydDatasheetsDatasetsHelp2021}
Boyd, K.~L. 2021.
\newblock Datasheets for {Datasets} help {ML} {Engineers} {Notice} and {Understand} {Ethical} {Issues} in {Training} {Data}.
\newblock \emph{Proceedings of the ACM on Human-Computer Interaction}, 5(CSCW2): 1--27.

\bibitem[{Brawley and Pury(2016)}]{brawleyWorkExperiencesMTurk2016}
Brawley, A.~M.; and Pury, C.~L. 2016.
\newblock Work experiences on {MTurk}: {Job} satisfaction, turnover, and information sharing.
\newblock \emph{Computers in Human Behavior}, 54: 531--546.

\bibitem[{Buolamwini and Gebru(2018)}]{buolamwiniGenderShadesIntersectional2018}
Buolamwini, J.; and Gebru, T. 2018.
\newblock Gender {Shades}: {Intersectional} {Accuracy} {Disparities} in {Commercial} {Gender} {Classification}.
\newblock In Friedler, S.~A.; and Wilson, C., eds., \emph{Proceedings of the 1st {Conference} on {Fairness}, {Accountability} and {Transparency}}, volume~81, 77--91. Proceedings of Machine Learning Research: PMLR.

\bibitem[{Chasalow and Levy(2021)}]{chasalowRepresentativenessStatisticsPolitics2021}
Chasalow, K.; and Levy, K. 2021.
\newblock Representativeness in Statistics, Politics, and Machine Learning.
\newblock In \emph{Proceedings of the 2021 ACM Conference on Fairness, Accountability, and Transparency}, FAccT ’21, 77–89. New York, NY, USA: Association for Computing Machinery.
\newblock ISBN 978-1-4503-8309-7.

\bibitem[{Cherry(2016)}]{cherry2016virtual}
Cherry, M.~A. 2016.
\newblock Virtual work and invisible labor.
\newblock \emph{Invisible labour: Hidden work in the contemporary world}, 28--46.

\bibitem[{Conner-Simons(2021)}]{MajorMLDatasetsa}
Conner-Simons, A. 2021.
\newblock Major ML datasets have tens of thousands of errors | MIT CSAIL.
\newblock \emph{MIT CSAIL News}.
\newblock Citation Key: MajorMLDatasetsa.

\bibitem[{Couldry and Mejias(2019)}]{couldryDataColonialismRethinking2019}
Couldry, N.; and Mejias, U.~A. 2019.
\newblock Data {Colonialism}: {Rethinking} {Big} {Data}’s {Relation} to the {Contemporary} {Subject}.
\newblock \emph{Television \& New Media}, 20(4): 336--349.

\bibitem[{Coworker.org()}]{ENDHARMMASS2023}
Coworker.org. 2023.
\newblock {END} {THE} {HARM} {OF} {MASS} {REJECTIONS}.

\bibitem[{Crain, Poster, and Cherry(2016)}]{crainInvisibleLaborHidden2016}
Crain, M.~G.; Poster, W.~R.; and Cherry, M.~A., eds. 2016.
\newblock \emph{Invisible {Labor}: {Hidden} {Work} in the {Contemporary} {World}}.
\newblock University of California Press, 1 edition.
\newblock ISBN 978-0-520-28640-5.

\bibitem[{Denton et~al.(2021{\natexlab{a}})Denton, D{\'\i}az, Kivlichan, Prabhakaran, and Rosen}]{denton2021whose}
Denton, E.; D{\'\i}az, M.; Kivlichan, I.; Prabhakaran, V.; and Rosen, R. 2021{\natexlab{a}}.
\newblock Whose ground truth? accounting for individual and collective identities underlying dataset annotation.
\newblock \emph{arXiv preprint arXiv:2112.04554}.

\bibitem[{Denton et~al.(2021{\natexlab{b}})Denton, Hanna, Amironesei, Smart, and Nicole}]{dentonGenealogyMachineLearning2021}
Denton, E.; Hanna, A.; Amironesei, R.; Smart, A.; and Nicole, H. 2021{\natexlab{b}}.
\newblock On the genealogy of machine learning datasets: {A} critical history of {ImageNet}.
\newblock \emph{Big Data \& Society}, 8(2): 205395172110359.

\bibitem[{D\'{\i}az et~al.(2022)D\'{\i}az, Kivlichan, Rosen, Baker, Amironesei, Prabhakaran, and Denton}]{CrowdSheets}
D\'{\i}az, M.; Kivlichan, I.; Rosen, R.; Baker, D.; Amironesei, R.; Prabhakaran, V.; and Denton, E. 2022.
\newblock CrowdWorkSheets: Accounting for Individual and Collective Identities Underlying Crowdsourced Dataset Annotation.
\newblock In \emph{2022 ACM Conference on Fairness, Accountability, and Transparency}, FAccT '22, 2342–2351. New York, NY, USA: Association for Computing Machinery.
\newblock ISBN 9781450393522.

\bibitem[{D'Ignazio and Klein(2020)}]{dignazioDataFeminism2020}
D'Ignazio, C.; and Klein, L.~F. 2020.
\newblock \emph{Data feminism}.
\newblock Strong ideas series. Cambridge, Massachusetts: The MIT Press.
\newblock ISBN 978-0-262-04400-4.

\bibitem[{Edwards and Holland(2013)}]{edwardsWhatQualitativeInterviewing2013}
Edwards, R.; and Holland, J. 2013.
\newblock \emph{What is qualitative interviewing?}
\newblock What is? {Research} methods series. London : New Delhi: Bloomsbury.
\newblock ISBN 978-1-78093-852-3 978-1-84966-809-5.
\newblock OCLC: ocn855705441.

\bibitem[{Espeland and Stevens(2008)}]{Espeland_Stevens_2008}
Espeland, W.~N.; and Stevens, M.~L. 2008.
\newblock A Sociology of Quantification.
\newblock \emph{European Journal of Sociology / Archives Européennes de Sociologie}, 49(3): 401–436.

\bibitem[{Feinberg(2017)}]{feinbergDesignPerspectiveData2017}
Feinberg, M. 2017.
\newblock A {Design} {Perspective} on {Data}.
\newblock In \emph{Proceedings of the 2017 {CHI} {Conference} on {Human} {Factors} in {Computing} {Systems}}, 2952--2963. Denver Colorado USA: ACM.
\newblock ISBN 978-1-4503-4655-9.

\bibitem[{Fieseler, Bucher, and Hoffmann(2019)}]{fieselerUnfairnessDesignPerceived2019}
Fieseler, C.; Bucher, E.; and Hoffmann, C.~P. 2019.
\newblock Unfairness by {Design}? {The} {Perceived} {Fairness} of {Digital} {Labor} on {Crowdworking} {Platforms}.
\newblock \emph{Journal of Business Ethics}, 156(4): 987--1005.

\bibitem[{Flick(2018)}]{flickDoingGroundedTheory2018}
Flick, U. 2018.
\newblock \emph{Doing grounded theory}.
\newblock Number 8. volume in The {SAGE} qualitative research kit / edited by {Uwe} {Flick}. Los Angeles London New Dehli Singapore Washington DC Melbourne: SAGE, 2nd edition edition.
\newblock ISBN 978-1-4739-1200-7.

\bibitem[{Gebru et~al.(2021)Gebru, Morgenstern, Vecchione, Vaughan, Wallach, Iii, and Crawford}]{gebruDatasheetsDatasets2021}
Gebru, T.; Morgenstern, J.; Vecchione, B.; Vaughan, J.~W.; Wallach, H.; Iii, H.~D.; and Crawford, K. 2021.
\newblock Datasheets for datasets.
\newblock \emph{Communications of the ACM}, 64(12): 86--92.

\bibitem[{Gerchick et~al.(2023)Gerchick, Jegede, Shah, Gutierrez, Beiers, Shemtov, Xu, Samant, and Horowitz}]{gerchickDevilDetailsInterrogating2023}
Gerchick, M.; Jegede, T.; Shah, T.; Gutierrez, A.; Beiers, S.; Shemtov, N.; Xu, K.; Samant, A.; and Horowitz, A. 2023.
\newblock The {Devil} is in the {Details}: {Interrogating} {Values} {Embedded} in the {Allegheny} {Family} {Screening} {Tool}.
\newblock In \emph{Proceedings of the 2023 {ACM} {Conference} on {Fairness}, {Accountability}, and {Transparency}}, {FAccT} '23, 1292--1310. New York, NY, USA: Association for Computing Machinery.
\newblock ISBN 9798400701924.

\bibitem[{Gitelman(2013)}]{gitelmanRawDataOxymoron2013}
Gitelman, L., ed. 2013.
\newblock \emph{"{Raw} data" is an oxymoron}.
\newblock Infrastructures series. Cambridge, Massachusetts ; London, England: The MIT Press.
\newblock ISBN 978-0-262-51828-4.

\bibitem[{Gray and Suri(2019)}]{grayGhostWorkHow2019}
Gray, M.~L.; and Suri, S. 2019.
\newblock \emph{Ghost work: how to stop {Silicon} {Valley} from building a new global underclass}.
\newblock Boston: Houghton Mifflin Harcourt.
\newblock ISBN 978-1-328-56628-7.

\bibitem[{Hanrahan et~al.(2018)Hanrahan, Martin, Willamowski, and Carroll}]{hanrahanInvestigatingAmazonMechanical2018}
Hanrahan, B.~V.; Martin, D.; Willamowski, J.; and Carroll, J.~M. 2018.
\newblock Investigating the {Amazon} {Mechanical} {Turk} {Market} {Through} {Tool} {Design}.
\newblock \emph{Computer Supported Cooperative Work (CSCW)}, 27(3-6): 1255--1274.

\bibitem[{Hasselbalch(2021)}]{Hasselbalch_2021}
Hasselbalch, G. 2021.
\newblock \emph{Data Ethics of Power: A Human Approach in the Big Data and AI Era}.
\newblock Edward Elgar Publishing.
\newblock ISBN 978-1-80220-311-0.

\bibitem[{Hawkins and Mittelstadt(2023)}]{hawkinsEthicalAmbiguityAI2023}
Hawkins, W.; and Mittelstadt, B. 2023.
\newblock The ethical ambiguity of {AI} data enrichment: {Measuring} gaps in research ethics norms and practices.
\newblock In \emph{Proceedings of the 2023 {ACM} {Conference} on {Fairness}, {Accountability}, and {Transparency}}, {FAccT} '23, 261--270. New York, NY, USA: Association for Computing Machinery.
\newblock ISBN 9798400701924.

\bibitem[{Hitlin(2016)}]{hitlin_research_2016}
Hitlin, P. 2016.
\newblock Research in the {Crowdsourcing} {Age}, a {Case} {Study}.
\newblock Technical report, Pew Research Center.

\bibitem[{Howe(1978)}]{howePinkCollarWorkers1978}
Howe, L.~K. 1978.
\newblock \emph{Pink {Collar} {Workers}: {Inside} the {World} of {Women}'s {Work}}.
\newblock Avon.
\newblock ISBN 978-0-380-01924-3.
\newblock Google-Books-ID: 10K7AAAAIAAJ.

\bibitem[{Irani and Silberman(2013)}]{iraniTurkopticonInterruptingWorker2013}
Irani, L.~C.; and Silberman, M.~S. 2013.
\newblock Turkopticon: interrupting worker invisibility in amazon mechanical turk.
\newblock In \emph{Proceedings of the {SIGCHI} {Conference} on {Human} {Factors} in {Computing} {Systems}}, 611--620. Paris France: ACM.
\newblock ISBN 978-1-4503-1899-0.

\bibitem[{Irani and Silberman(2016)}]{iraniStoriesWeTell2016}
Irani, L.~C.; and Silberman, M.~S. 2016.
\newblock Stories {We} {Tell} {About} {Labor}: {Turkopticon} and the {Trouble} with "{Design}".
\newblock In \emph{Proceedings of the 2016 {CHI} {Conference} on {Human} {Factors} in {Computing} {Systems}}, 4573--4586. San Jose California USA: ACM.
\newblock ISBN 978-1-4503-3362-7.

\bibitem[{Jacobs and Wallach(2021)}]{Jacobs_Wallach_2021}
Jacobs, A.~Z.; and Wallach, H. 2021.
\newblock Measurement and Fairness.
\newblock In \emph{Proceedings of the 2021 ACM Conference on Fairness, Accountability, and Transparency}, FAccT ’21, 375–385. New York, NY, USA: Association for Computing Machinery.
\newblock ISBN 978-1-4503-8309-7.

\bibitem[{Jo and Gebru(2020)}]{jo2020lessons}
Jo, E.~S.; and Gebru, T. 2020.
\newblock Lessons from Archives: Strategies for Collecting Sociocultural Data in Machine Learning.
\newblock In \emph{Proceedings of the 2020 Conference on Fairness, Accountability, and Transparency}, FAT* '20, 306–316. New York, NY, USA: Association for Computing Machinery.
\newblock ISBN 9781450369367.

\bibitem[{Jones(2021)}]{jonesWorkWorkerLabour2021}
Jones, P. 2021.
\newblock \emph{Work without the worker: labour in the age of platform capitalism}.
\newblock Brooklyn: Verso Books.
\newblock ISBN 978-1-83976-043-3.

\bibitem[{Kapania, Taylor, and Wang(2023)}]{kapaniaHuntSnarkAnnotator2023}
Kapania, S.; Taylor, A.~S.; and Wang, D. 2023.
\newblock A hunt for the {Snark}: {Annotator} {Diversity} in {Data} {Practices}.
\newblock In \emph{Proceedings of the 2023 {CHI} {Conference} on {Human} {Factors} in {Computing} {Systems}}, 1--15. Hamburg Germany: ACM.
\newblock ISBN 978-1-4503-9421-5.

\bibitem[{Leavy, Siapera, and O'Sullivan(2021)}]{leavyEthicalDataCuration2021}
Leavy, S.; Siapera, E.; and O'Sullivan, B. 2021.
\newblock Ethical {Data} {Curation} for {AI}: {An} {Approach} based on {Feminist} {Epistemology} and {Critical} {Theories} of {Race}.
\newblock In \emph{Proceedings of the 2021 {AAAI}/{ACM} {Conference} on {AI}, {Ethics}, and {Society}}, 695--703. Virtual Event USA: ACM.
\newblock ISBN 978-1-4503-8473-5.

\bibitem[{Li et~al.(2023)Li, Vincent, Chancellor, and Hecht}]{liDimensionsDataLabor2023a}
Li, H.; Vincent, N.; Chancellor, S.; and Hecht, B. 2023.
\newblock The {Dimensions} of {Data} {Labor}: {A} {Road} {Map} for {Researchers}, {Activists}, and {Policymakers} to {Empower} {Data} {Producers}.
\newblock In \emph{Proceedings of the 2023 {ACM} {Conference} on {Fairness}, {Accountability}, and {Transparency}}, {FAccT} '23, 1151--1161. New York, NY, USA: Association for Computing Machinery.
\newblock ISBN 9798400701924.

\bibitem[{Maund et~al.(2020)Maund, Irvine, Lawson, Steadman, Risely, Cunningham, and Davies}]{maund2020motivates}
Maund, P.~R.; Irvine, K.~N.; Lawson, B.; Steadman, J.; Risely, K.; Cunningham, A.~A.; and Davies, Z.~G. 2020.
\newblock What motivates the masses: Understanding why people contribute to conservation citizen science projects.
\newblock \emph{Biological Conservation}, 246: 108587.

\bibitem[{Mayer-Schönberger and Cukier(2014)}]{Mayer-Schönberger_Cukier_2014}
Mayer-Schönberger, V.; and Cukier, K. 2014.
\newblock \emph{Big data: a revolution that will transform how we live, work, and think}.
\newblock Boston: Mariner Books, Houghton Mifflin Harcourt, first mariner books edition edition.
\newblock ISBN 978-0-544-22775-0.

\bibitem[{Miceli and Posada(2022)}]{miceliDataProductionDispositif2022}
Miceli, M.; and Posada, J. 2022.
\newblock The {Data}-{Production} {Dispositif}.
\newblock \emph{Proceedings of the ACM on Human-Computer Interaction}, 6(CSCW2): 460:1--460:37.

\bibitem[{Miceli, Posada, and Yang(2022)}]{miceliStudyingMachineLearning2022}
Miceli, M.; Posada, J.; and Yang, T. 2022.
\newblock Studying {Up} {Machine} {Learning} {Data}: {Why} {Talk} {About} {Bias} {When} {We} {Mean} {Power}?
\newblock \emph{Proceedings of the ACM on Human-Computer Interaction}, 6(GROUP): 1--14.

\bibitem[{Miceli, Schuessler, and Yang(2020)}]{miceli2020between}
Miceli, M.; Schuessler, M.; and Yang, T. 2020.
\newblock Between subjectivity and imposition: Power dynamics in data annotation for computer vision.
\newblock \emph{Proceedings of the ACM on Human-Computer Interaction}, 4(CSCW2): 1--25.

\bibitem[{Miceli et~al.(2022)Miceli, Yang, Alvarado~Garcia, Posada, Wang, Pohl, and Hanna}]{Miceli_Yang_AlvaradoGarcia_Posada_Wang_Pohl_Hanna_2022}
Miceli, M.; Yang, T.; Alvarado~Garcia, A.; Posada, J.; Wang, S.~M.; Pohl, M.; and Hanna, A. 2022.
\newblock Documenting Data Production Processes: A Participatory Approach for Data Work.
\newblock \emph{Proceedings of the ACM on Human-Computer Interaction}, 6(CSCW2): 1–34.

\bibitem[{Mittelstadt, Wachter, and Russell(2023)}]{mittelstadtUnfairnessFairMachine2023}
Mittelstadt, B.; Wachter, S.; and Russell, C. 2023.
\newblock The {Unfairness} of {Fair} {Machine} {Learning}: {Levelling} down and strict egalitarianism by default.

\bibitem[{Muller et~al.(2019{\natexlab{a}})Muller, George, John, Passi, Feinberg, Jackson, and Kery}]{mullerHumanCenteredStudyData2019}
Muller, M.; George, T.; John, B.~E.; Passi, S.; Feinberg, M.; Jackson, S.~J.; and Kery, M.~B. 2019{\natexlab{a}}.
\newblock Human-Centered Study of Data Science Work Practices.
\newblock In \emph{Extended Abstracts of the 2019 CHI Conference}, 8.
\newblock Citation Key: mullerHumanCenteredStudyData2019.

\bibitem[{Muller et~al.(2019{\natexlab{b}})Muller, Lange, Wang, Piorkowski, Tsay, Liao, Dugan, and Erickson}]{mullerHowDataScience2019a}
Muller, M.; Lange, I.; Wang, D.; Piorkowski, D.; Tsay, J.; Liao, Q.~V.; Dugan, C.; and Erickson, T. 2019{\natexlab{b}}.
\newblock How {Data} {Science} {Workers} {Work} with {Data}: {Discovery}, {Capture}, {Curation}, {Design}, {Creation}.
\newblock In \emph{Proceedings of the 2019 {CHI} {Conference} on {Human} {Factors} in {Computing} {Systems}}, 1--15. Glasgow Scotland Uk: ACM.
\newblock ISBN 978-1-4503-5970-2.

\bibitem[{Njanja(2022)}]{njanjaMetaBarredKenyan2023}
Njanja, A. 2022.
\newblock Meta sued by Ethiopians and Kenyan rights group for fueling Tigray War.
\newblock \emph{TechCrunch}.
\newblock Citation Key: njanjaMetaSuedEthiopians2022.

\bibitem[{OSTP(2022)}]{OSTP_2022}
OSTP. 2022.
\newblock Blueprint for an AI Bill of Rights.
\newblock Citation Key: BlueprintAIBill.

\bibitem[{Papakyriakopoulos et~al.(2023)Papakyriakopoulos, Choi, Thong, Zhao, Andrews, Bourke, Xiang, and Koenecke}]{papakyriakopoulosAugmentedDatasheetsSpeech2023}
Papakyriakopoulos, O.; Choi, A. S.~G.; Thong, W.; Zhao, D.; Andrews, J.; Bourke, R.; Xiang, A.; and Koenecke, A. 2023.
\newblock Augmented {Datasheets} for {Speech} {Datasets} and {Ethical} {Decision}-{Making}.
\newblock In \emph{Proceedings of the 2023 {ACM} {Conference} on {Fairness}, {Accountability}, and {Transparency}}, {FAccT} '23, 881--904. New York, NY, USA: Association for Computing Machinery.
\newblock ISBN 9798400701924.

\bibitem[{Paullada et~al.(2021)Paullada, Raji, Bender, Denton, and Hanna}]{Paullada_Raji_Bender_Denton_Hanna_2021}
Paullada, A.; Raji, I.~D.; Bender, E.~M.; Denton, E.; and Hanna, A. 2021.
\newblock Data and its (dis)contents: A survey of dataset development and use in machine learning research.
\newblock \emph{Patterns}, 2(11): 100336.

\bibitem[{Perrigo(2022)}]{MetaAccusedHuman}
Perrigo, B. 2022.
\newblock Meta Accused Of Human Trafficking and Union-Busting in Kenya.
\newblock \emph{TIME}.

\bibitem[{Pontin(2007)}]{Pontin_2007}
Pontin, J. 2007.
\newblock Artificial Intelligence, With Help From the Humans.
\newblock \emph{The New York Times}.

\bibitem[{Qadri et~al.(2023)Qadri, Shelby, Bennett, and Denton}]{qadriAIRegimesRepresentation2023}
Qadri, R.; Shelby, R.; Bennett, C.~L.; and Denton, E. 2023.
\newblock {AI}’s {Regimes} of {Representation}: {A} {Community}-centered {Study} of {Text}-to-{Image} {Models} in {South} {Asia}.
\newblock In \emph{Proceedings of the 2023 {ACM} {Conference} on {Fairness}, {Accountability}, and {Transparency}}, {FAccT} '23, 506--517. New York, NY, USA: Association for Computing Machinery.
\newblock ISBN 9798400701924.

\bibitem[{Rolf et~al.(2021)Rolf, Worledge, Recht, and Jordan}]{rolf2021representation}
Rolf, E.; Worledge, T.~T.; Recht, B.; and Jordan, M. 2021.
\newblock Representation matters: Assessing the importance of subgroup allocations in training data.
\newblock In \emph{International Conference on Machine Learning}, 9040--9051. PMLR.

\bibitem[{Rostamzadeh et~al.(2022)Rostamzadeh, Mincu, Roy, Smart, Wilcox, Pushkarna, Schrouff, Amironesei, Moorosi, and Heller}]{Healthsheet}
Rostamzadeh, N.; Mincu, D.; Roy, S.; Smart, A.; Wilcox, L.; Pushkarna, M.; Schrouff, J.; Amironesei, R.; Moorosi, N.; and Heller, K. 2022.
\newblock Healthsheet: Development of a Transparency Artifact for Health Datasets.
\newblock In \emph{Proceedings of the 2022 ACM Conference on Fairness, Accountability, and Transparency}, FAccT '22, 1943–1961. New York, NY, USA: Association for Computing Machinery.
\newblock ISBN 9781450393522.

\bibitem[{Rothschild et~al.(2022)Rothschild, Booker, Davoll, Hill, Ivey, DiSalvo, Rydal~Shapiro, and DiSalvo}]{rothschildFairProsocialEmployment2022}
Rothschild, A.; Booker, J.; Davoll, C.; Hill, J.; Ivey, V.; DiSalvo, C.; Rydal~Shapiro, B.; and DiSalvo, B. 2022.
\newblock Towards fair and pro-social employment of digital pieceworkers for sourcing machine learning training data.
\newblock In \emph{Extended {Abstracts} of the 2022 {CHI} {Conference} on {Human} {Factors} in {Computing} {Systems}}, {CHI} {EA} '22, 1--9. New York, NY, USA: Association for Computing Machinery.
\newblock ISBN 978-1-4503-9156-6.

\bibitem[{Rowe(2023)}]{roweItDestroyedMe2023}
Rowe, N. 2023.
\newblock ‘{It}’s destroyed me completely’: {Kenyan} moderators decry toll of training of {AI} models.
\newblock \emph{The Guardian}.

\bibitem[{Sainato(2023)}]{Sainato_2023}
Sainato, M. 2023.
\newblock ‘War of attrition’: why union victories for US workers at Amazon have stalled.
\newblock \emph{The Guardian}.

\bibitem[{Salehi et~al.(2015)Salehi, Irani, Bernstein, Alkhatib, Ogbe, Milland, and {Clickhappier}}]{salehiWeAreDynamo2015}
Salehi, N.; Irani, L.~C.; Bernstein, M.~S.; Alkhatib, A.; Ogbe, E.; Milland, K.; and {Clickhappier}. 2015.
\newblock We {Are} {Dynamo}: {Overcoming} {Stalling} and {Friction} in {Collective} {Action} for {Crowd} {Workers}.
\newblock In \emph{Proceedings of the 33rd {Annual} {ACM} {Conference} on {Human} {Factors} in {Computing} {Systems}}, 1621--1630. Seoul Republic of Korea: ACM.
\newblock ISBN 978-1-4503-3145-6.

\bibitem[{Sambasivan et~al.(2021)Sambasivan, Kapania, Highfill, Akrong, Paritosh, and Aroyo}]{sambasivanEveryoneWantsModel2021b}
Sambasivan, N.; Kapania, S.; Highfill, H.; Akrong, D.; Paritosh, P.; and Aroyo, L.~M. 2021.
\newblock "{Everyone} wants to do the model work, not the data work": {Data} {Cascades} in {High}-{Stakes} {AI}.
\newblock In \emph{Proceedings of the 2021 {CHI} {Conference} on {Human} {Factors} in {Computing} {Systems}}, {CHI} '21, 1--15. New York, NY, USA: Association for Computing Machinery.
\newblock ISBN 978-1-4503-8096-6.

\bibitem[{Sambasivan and Veeraraghavan(2022)}]{sambasivanDeskillingDomainExpertise2022a}
Sambasivan, N.; and Veeraraghavan, R. 2022.
\newblock The {Deskilling} of {Domain} {Expertise} in {AI} {Development}.
\newblock In \emph{Proceedings of the 2022 {CHI} {Conference} on {Human} {Factors} in {Computing} {Systems}}, {CHI} '22, 1--14. New York, NY, USA: Association for Computing Machinery.
\newblock ISBN 978-1-4503-9157-3.

\bibitem[{Sannon and Cosley(2019)}]{sannonPrivacyPowerInvisible2019a}
Sannon, S.; and Cosley, D. 2019.
\newblock Privacy, {Power}, and {Invisible} {Labor} on {Amazon} {Mechanical} {Turk}.
\newblock In \emph{Proceedings of the 2019 {CHI} {Conference} on {Human} {Factors} in {Computing} {Systems}}, 1--12. Glasgow Scotland Uk: ACM.
\newblock ISBN 978-1-4503-5970-2.

\bibitem[{Scheuerman, Hanna, and Denton(2021)}]{scheuerman2021datasets}
Scheuerman, M.~K.; Hanna, A.; and Denton, E. 2021.
\newblock Do Datasets Have Politics? Disciplinary Values in Computer Vision Dataset Development.
\newblock \emph{Proc. ACM Hum.-Comput. Interact.}, 5(CSCW2).

\bibitem[{Schwartz(2019)}]{UntoldHistoryAI2019}
Schwartz, O. 2019.
\newblock Untold History of AI: How Amazon’s Mechanical Turkers Got Squeezed Inside the Machine.
\newblock \emph{IEEE Spectrum}.
\newblock Citation Key: UntoldHistoryAI2019.

\bibitem[{Srinivasan et~al.(2021)Srinivasan, Denton, Famularo, Rostamzadeh, Diaz, and Coleman}]{Artsheets}
Srinivasan, R.; Denton, E.; Famularo, J.; Rostamzadeh, N.; Diaz, F.; and Coleman, B. 2021.
\newblock Artsheets for art datasets.
\newblock In \emph{Thirty-fifth conference on neural information processing systems datasets and benchmarks track (round 2)}.

\bibitem[{Strauss and Corbin(1997)}]{straussGroundedTheoryPractice1997}
Strauss, A.~L.; and Corbin, J.~M., eds. 1997.
\newblock \emph{Grounded theory in practice}.
\newblock Thousand Oaks: Sage Publications.
\newblock ISBN 978-0-7619-0747-3 978-0-7619-0748-0.

\bibitem[{Thebault-Spieker et~al.(2023)Thebault-Spieker, Venkatagiri, Mine, and Luther}]{diverse-perspectives-can}
Thebault-Spieker, J.; Venkatagiri, S.; Mine, N.; and Luther, K. 2023.
\newblock Diverse Perspectives Can Mitigate Political Bias in Crowdsourced Content Moderation.
\newblock In \emph{Proceedings of the 2023 ACM Conference on Fairness, Accountability, and Transparency}, FAccT '23, 1280–1291. New York, NY, USA: Association for Computing Machinery.
\newblock ISBN 9798400701924.

\bibitem[{Thylstrup et~al.(2022)Thylstrup, Hansen, Flyverbom, and Amoore}]{Thylstrup_Hansen_Flyverbom_Amoore_2022}
Thylstrup, N.~B.; Hansen, K.~B.; Flyverbom, M.; and Amoore, L. 2022.
\newblock Politics of data reuse in machine learning systems: Theorizing reuse entanglements.
\newblock \emph{Big Data \& Society}, 9(2): 20539517221139785.

\bibitem[{Tubaro, Casilli, and Coville(2020)}]{Tubaro_Casilli_Coville_2020}
Tubaro, P.; Casilli, A.~A.; and Coville, M. 2020.
\newblock The trainer, the verifier, the imitator: Three ways in which human platform workers support artificial intelligence.
\newblock \emph{Big Data \& Society}, 7(1): 2053951720919776.

\bibitem[{Valdivia and Tazzioli(2023)}]{valdiviaDataficationGenealogiesAlgorithmic2023}
Valdivia, A.; and Tazzioli, M. 2023.
\newblock Datafication {Genealogies} beyond {Algorithmic} {Fairness}: {Making} {Up} {Racialised} {Subjects}.
\newblock In \emph{Proceedings of the 2023 {ACM} {Conference} on {Fairness}, {Accountability}, and {Transparency}}, {FAccT} '23, 840--850. New York, NY, USA: Association for Computing Machinery.
\newblock ISBN 9798400701924.

\bibitem[{Vasudevan et~al.(2022)Vasudevan, Caine, Gontijo~Lopes, Fridovich-Keil, and Roelofs}]{vasudevan_when_2022}
Vasudevan, V.; Caine, B.; Gontijo~Lopes, R.; Fridovich-Keil, S.; and Roelofs, R. 2022.
\newblock When does dough become a bagel? Analyzing the remaining mistakes on ImageNet.
\newblock In Koyejo, S.; Mohamed, S.; Agarwal, A.; Belgrave, D.; Cho, K.; and Oh, A., eds., \emph{Advances in neural information processing systems}, volume~35, 6720–6734. Curran Associates, Inc.

\bibitem[{Vertesi(2014)}]{vertesiSeamfulSpacesHeterogeneous2014}
Vertesi, J. 2014.
\newblock Seamful {Spaces}: {Heterogeneous} {Infrastructures} in {Interaction}.
\newblock \emph{Science, Technology, \& Human Values}, 39(2): 264--284.

\bibitem[{Wang, Prabhat, and Sambasivan(2022)}]{wang2022whose}
Wang, D.; Prabhat, S.; and Sambasivan, N. 2022.
\newblock Whose AI Dream? In search of the aspiration in data annotation.
\newblock In \emph{Proceedings of the 2022 CHI Conference on Human Factors in Computing Systems}, 1--16.

\bibitem[{Wikipedia. ``Blue-collar work''()}]{Blue-collarworker_2024}
Wikipedia. ``Blue-collar work''. 2024.

\bibitem[{Wikipedia. ``White-collar work''()}]{White-collar-worker_2024}
Wikipedia. ``White-collar work''. 2024.

\bibitem[{Xia et~al.(2017)Xia, Wang, Huang, and Shah}]{xiaOurPrivacyNeeds2017}
Xia, H.; Wang, Y.; Huang, Y.; and Shah, A. 2017.
\newblock "{Our} {Privacy} {Needs} to be {Protected} at {All} {Costs}": {Crowd} {Workers}' {Privacy} {Experiences} on {Amazon} {Mechanical} {Turk}.
\newblock \emph{Proceedings of the ACM on Human-Computer Interaction}, 1(CSCW): 1--22.

\end{thebibliography}

\section{Appendix}
For an overview of the proxies employed by requesters, see Table \ref{tab:proxies} (next page).


\begin{center}
    
\begin{table}
    \centering
    \captionsetup{justification=centering}
    \caption{Proxies user by requesters to approximate worker identity or aptitude}
\label{tab:proxies}
    \begin{tabular}{| p{3cm} | p{4cm} | p{5cm} |}
        \hline
        \textbf{Category} & \textbf{Sub-category} & \textbf{Description} \\
        \hline
        Worker identity & English fluency & Worker English fluency or skill level \\
        \cline{2-3}
         & Age & Worker legal age \\
        \cline{2-3}
         & Location and time zone & Worker IP address and time zone as signal of physical location \\
        \hline
        Worker aptitude & Pre hoc: Approval rating & Aggregate approval rating for prior tasks completed on platform, as rated by requester \\
        \cline{2-3}
         & Pre hoc: Number of tasks completed & Total number prior tasks completed on the platform \\
        \cline{2-3}
         & Post hoc: Keyboard interaction & Speed of typing, calculating amount of text copied and pasted \\
        \cline{2-3}
         & Post hoc: Answer pattern & On sequential multiple choice and Likert scale questions, visual pattern (whether or not uniform) of answer selection between questions \\
        \cline{2-3}
         & Post hoc: Attention check & Requiring human-only review (i.e., catches bots) and/or misleading questions meant to trip up non-careful readers \\
        \cline{2-3}
         & Post hoc: Gut reaction & An instinctual feeling of whether or not requester felt task done correctly \\
        \cline{2-3}
         & Post hoc: Coherence as trust & Assuming workers who answered consistently with majority choice performed work sufficiently; or worker submission matched ``gold standard'' label \\
        \hline
    \end{tabular}
    
\end{table}
\end{center}
\end{document}